\documentclass[aps,prd,preprint,showpacs]{revtex4}

\usepackage{graphicx}
\usepackage{bm}
\makeatother
\begin{document}

\title{Improving the Lagrangian perturbative solution for cosmic fluid:
 Applying Shanks transformation}
\author{Takayuki Tatekawa$^{1, 2}$\thanks{E-mail: tatekawa@cpd.kogakuin.ac.jp}
 }

\affiliation{
1. The Center for Continuing Professional Development,
Kogakuin University,
1-24-2 Nishi-shinjuku, Shinjuku, Tokyo 163-8677 Japan\\
2. Advanced Research Institute for Science
and Engineering, Waseda University,
3-4-1 Okubo, Shinjuku, Tokyo 169-8555, Japan}

\date{\today}
%%%%%%%%%%%%%%%%%%%%%%%%%%%%%%%%%%%%%%%%%%%%%%%
\begin{abstract}
We study the behavior of Lagrangian perturbative solutions.
For a spherical void model, the higher order
the Lagrangian perturbation we consider, the worse the
approximation becomes in late-time evolution.
In particular, if we stop to improve until an even order is reached,
the perturbative solution describes the contraction of the void.
To solve this problem,
we consider improving the perturbative solution using
Shanks transformation, which accelerates
the convergence of the sequence. After the transformation,
we find that the accuracy of higher-order perturbation
is recovered and the perturbative solution is refined well.
Then we show that this improvement method can apply
for a $\Lambda$CDM model and improved the power spectrum
of the density field.
\end{abstract}
%%%%%%%%%%%%%%%%%%%%%%%%%%%%%%%%%%%%%%%%%%%%%%%


\maketitle

%%%%%%%%%%%%%%%%%%%%%%%%%%%%%%%%%%%%%%%%%%%
\section{Introduction}\label{sec:intro}
%%%%%%%%%%%%%%%%%%%%%%%%%%%%%%%%%%%%%%%%%%%
There are various structures in the universe
which are gravitationally bounded, for example,
galaxies, groups of galaxies, clusters of galaxies, 
voids, large-scale structure, and so on.
These structures have evolved spontaneously from 
a primordial density fluctuation.

The scenario for the growth of density perturbation is analyzed
by several methods.
When we do not consider the superhorizon scale or an extremely
dense region like a supermassive black hole,
the motion of the cosmological fluid can be described
by Newtonian cosmology. Further,
the Lagrangian description for the cosmological fluid
can be usefully applied to
the structure formation scenario. This description provides a
relatively accurate model even in a quasi-linear regime.
Zel'dovich~\cite{Zeldovich70} proposed
a linear Lagrangian approximation for dust fluid.
This approximation is called the Zel'dovich approximation
(ZA)~\cite{Zeldovich70,Shandarin89,Buchert89,Paddy93,Coles95,Sahni95,Jones05,Tatekawa04R,Paddy05}.
After that, higher-order approximation for the Lagrangian description
was proposed
\cite{Bouchet92,Buchert93,Buchert94,Bouchet95,Catelan95,Buchert92,Barrow93,Sasaki98}.

How accurate is the Lagrangian perturbation? To verify
its accuracy, we often use simple models to compare exact and
perturbative solutions.
One of the simplest models is the ``top-hat'' spherical symmetric
model, which has a constant density. For this model, we have obtained
an exact solution. Therefore, to estimate validity in
some approximated model, we often use the top-hat model.
According to recent analyses of several symmetric
models~\cite{Yoshisato06}, the spherical symmetric model has a 
little difficulty accurately describing evolution with the Lagrangian 
perturbation.

Munshi, Sahni, and Starobinsky~\cite{Munshi94} derived
up to the third-order perturbative solution.
In addition to these,
Sahni and Shandarin~\cite{Sahni96} obtained up to the
fifth-order perturbative solution.
We~\cite{Tatekawa04R} have derived up to the eleventh-order
solution.

If the density fluctuation is positive, the model will collapse. After
a caustic formation at the center of the model, the equation of
motion cannot describe the evolution.
In past analyses with the Lagrangian perturbation, if we consider
higher perturbative solutions, the approximation
is improved all the more. On the other hand, 
if the density fluctuation is negative, the spherical void expands.
In this case, ZA remains the best approximation to
apply to the late-time evolution of voids. Especially
if we stop expansion until an even order (2nd, 4th, 6th, $\cdots$)
is reached,
the perturbative solution describes the contraction of a void.
From the viewpoint of the convergence of the series,
we conjectured that the higher the order of perturbation
we consider, the worse the approximation becomes in
late-time evolution.

In this paper, we apply Shanks transformation, which accelerates
the convergence of the sequence~\cite{Sponier}.
When we regard the perturbative
solution as partial sums of infinite sequence, we must consider
sequence convergence if description accuracy is to be discussed.
Applying Shanks transformation,
we can converge the sequence with a few terms. Therefore,
we can improve the perturbative solutions.
Here we consider the spherical void evolution using Shanks
transformation. As a result, the higher-order perturbative
solution recovers its accuracy and describes late-time
evolution well. Because this method can apply not only
to a spherical model but also in a more generic case,
we can think that a new perturbative approach for 
Lagrangian description has been found.
From comparison of Shanks transformation and
Pad\'{e} approximation, which is another improvement method
for the sequence, we show several merits of
Shanks transformation for the perturbative approach.

To generalize our approach, we apply Shanks transformation
for a $\Lambda$CDM model. From the power spectrum of
the density field, we show that Shanks transformation also
recovers the accuracy of Lagrangian perturbation in a
$\Lambda$CDM model.

In Sec.~\ref{sec:evolution}, we briefly show the evolution equation
for a spherical symmetric model and derive exact solutions.
Then we introduce Lagrangian perturbative solutions (Sec.~\ref{sec:Lagrange}).
In Sec.~\ref{sec:Shanks}, we describe Shanks transformation, the
important method we have applied. Using this transformation, we
indicate the accuracy of new perturbative solutions and show
that Shanks transformation improves their solutions (Sec.~\ref{sec:example}).
For comparison, we also compute Pad\'{e} approximation
and show its behavior (Sec.~\ref{sec:Pade}).
In Sec.~\ref{sec:LCDM}, we consider the generic case and
the evolution of a $\Lambda$CDM model with N-body simulation and
Lagrangian approximations. Then we compare the power spectrum
of the density field
between the simulation and the approximations.
Finally we offer our summary and conclusion (Sec.~\ref{sec:summary}).

%%%%%%%%%%%%%%%%%%%%%%%%%%%%%%%%%%%%%%%%%%%
\section{Evolution equation and exact solutions}\label{sec:evolution}
%%%%%%%%%%%%%%%%%%%%%%%%%%%%%%%%%%%%%%%%%%%

We consider the ``top-hat'' spherical symmetric
model, which has a constant density.
In the E-dS Universe model,
the equation of motion of a spherical shell is written as
\begin{equation} \label{eqn:spherical-void}
\frac{\rm{d}}{{\rm d} t} \left (a^2 \frac{{\rm d} x}{{\rm d} t} \right )
= -\frac{2a^2 x}{9t^2} \left [ \left (\frac{x_0}{x}\right )^3 -1
 \right ] \,,
\end{equation}
where $x$ is a comoving radial coordinate and $x_0=x(t_0)$~\cite{Munshi94}.
Under the initial condition $|\delta|=a$ for $a \rightarrow 0$,
Eq.~(\ref{eqn:spherical-void}) can be integrated.
\begin{equation} \label{eqn:spherical-void2}
\left (\frac{{\rm d} R}{{\rm d} a} \right )^2 = a
 \left ( \frac{1}{R} - \frac{3}{5} \right ) \,,
\end{equation}
where $R(\theta) =a(t) x/x_0$ is a physical particle trajectory.
The exact solution for the spherical collapse
(Eq.~(\ref{eqn:spherical-void2})) can be parameterized as follows:
\begin{eqnarray}
R_+ (\theta) &=& \frac{3}{10} \left (1- \cos \theta \right ) \,, \\
a(\theta) &=& \frac{3}{5} \left [ \frac{3}{4}
 \left ( \theta - \sin \theta \right ) \right ]^{2/3} \,.
\end{eqnarray}
Similarly, the exact solution for the expansion of a top-hat void
(Eq.~(\ref{eqn:spherical-void2})) can be parameterized as follows:
\begin{eqnarray}
R_- (\theta) &=& \frac{3}{10} \left (\cosh \theta -1 \right ) \,, \\
a(\theta) &=& \frac{3}{5} \left [ \frac{3}{4}
 \left (\sinh \theta - \theta \right ) \right ]^{2/3} \,.
\end{eqnarray}
From these equations, we can obtain density fluctuation.
\begin{equation}
\delta (x) = \left ( \frac{x_0}{x} \right )^3 -1 \,,
\end{equation}
\begin{eqnarray}
\delta_+ (x)
 &=& \frac{9 (\theta-\sin \theta)^2}{2 (1 - \cos \theta)^3}
 -1 \,, \\
\delta_- (x)
 &=& \frac{9 (\theta-\sinh \theta)^2}{2 (\cosh \theta -1)^3}
 -1 \,,
\end{eqnarray}
where subscript $+$ and $-$ denote the case of spherical
collapse and of void expansion, respectively.

%%%%%%%%%%%%%%%%%%%%%%%%%%%%%%%%%%%%%%%%%%%
\section{Lagrangian perturbation}\label{sec:Lagrange}
%%%%%%%%%%%%%%%%%%%%%%%%%%%%%%%%%%%%%%%%%%%

In the Lagrangian description, the inhomogeneity of mass distribution
is described by the displacement from homogeneous distribution.
The Lagrangian perturbative solution for spherical symmetric models
in the E-dS Universe model is given by
\begin{equation} \label{eqn:Lagrange-void}
R_{\pm} (t)=R_0 \left [1- \sum_{k=1}^n (\pm 1)^{k} C_k a^k \right ] \,,
\end{equation}
where $C_k$ are Lagrangian perturbative coefficients.
The sign in coefficients corresponds to positive and negative density
fluctuation, respectively.
Substituting Eq.~(\ref{eqn:Lagrange-void}) to (\ref{eqn:spherical-void2}),
we derive the coefficients $C_k$.

Munshi, Sahni, and Starobinsky~\cite{Munshi94} derived
up to the third-order perturbative solution ($C_1, C_2, C_3$).
In addition to these,
Sahni and Shandarin~\cite{Sahni96} obtained $C_4$ and $C_5$.
Furthermore we~\cite{Tatekawa04R} derived $C_6, \cdots, C_{11}$.
The coefficients $C_k$ are shown in Table~\ref{tab:Lagrange}.

\begin{table}
\caption{\label{tab:Lagrange}
The perturbative coefficients in Lagrangian description.}
\vspace{0.3cm}
\begin{tabular}{l|c} \hline \hline
$k$ & $C_k$ \\ \hline
$1$ & $1/3$ \\
$2$ & $1/21$ \\
$3$ & $23/1701$ \\
$4$ & $1894/392931$ \\
$5$ & $3293/1702701$ \\
$6$ & $2418902/2896294401$ \\
$7$ & $55964945/147711014451$ \\
$8$ & $611605097/3430178002251$ \\
$9$ & $4529700278678/52512595036460559$ \\
$10$ & $2008868248800940/47103797747705121423$ \\
$11$ & $29117328566723/1356899523596443827$ \\ \hline
\end{tabular}
\end{table}

\begin{figure}
 \includegraphics{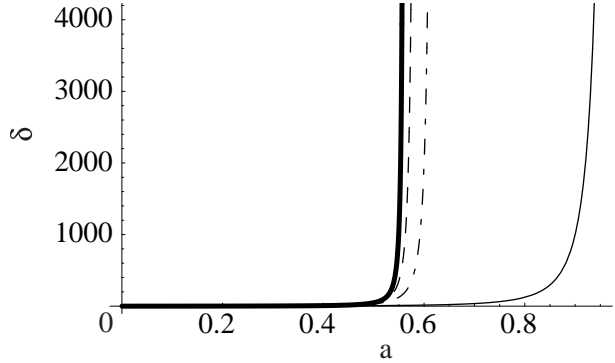}
 \caption{\label{fig:Cluster}
 The evolution of the spherical model for positive 
 fluctuation. The thick solid line shows evolution by exact solution.
 The fine solid line, the dashed-dotted line, and the dashed line
 show evolution by first-, fifth-, and eleventh-order
 Lagrangian perturbation, respectively. In this case, 
 the higher-order Lagrangian approximation
 gives an accurate description.
 In this figure, the scale factor is normalized by
 the time of a caustic formation in a first-order
 perturbation.
 }
\end{figure}

\begin{figure}
 \includegraphics{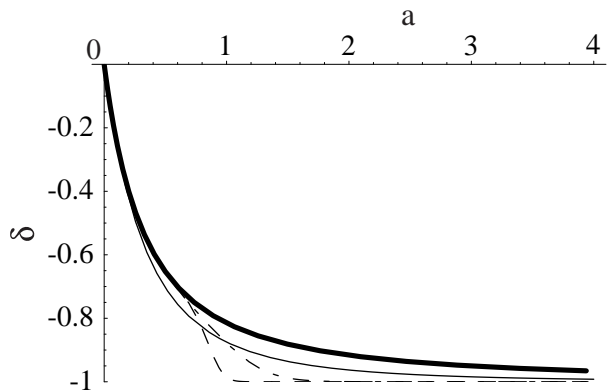}
 \caption{\label{fig:Void1}
 The evolution of the spherical model for negative 
 fluctuation.
 In this figure, the scale factor is normalized
 as in Fig.~\ref{fig:Cluster}.
 The thick solid line shows the evolution by the exact solution.
 The fine solid line, the dashed-dotted line, and the dashed line
 shows the evolution by first-, fifth-, and eleventh-order
 Lagrangian perturbation, respectively. In this case, 
 the higher-order Lagrangian approximation
 deviates from the exact solution at a late time.
 In other words, ZA gives the best description for the late-time
 evolution of a void.
 }
\end{figure}

The Lagrangian perturbation causes a serious problem.
If the density fluctuation is positive, the spherical fluctuation
collapses. The higher-order Lagrangian approximation
gives accurate description (Fig.~\ref{fig:Cluster}).
On the other hand, if the density fluctuation is negative,
the spherical void expands.
In this case, ZA remains the best approximation to
apply to the late-time evolution of voids. Especially,
if we stop to improve until an even order (2nd, 4th, 6th, $\cdots$),
the perturbative solution describes the contraction of a void.
In other words, ZA gives the best description for the late-time
evolution of voids~(Fig.~\ref{fig:Void1}). 

%%% Figure %%%
%%%%%%%%%%%%%

%%% Figure %%%
%%%%%%%%%%%%%

%%%%%%%%%%%%%%%%%%%%%%%%%%%%%%%%%%%%%%%%%%%
\section{Shanks transformation}\label{sec:Shanks}
%%%%%%%%%%%%%%%%%%%%%%%%%%%%%%%%%%%%%%%%%%%

The Lagrangian perturbation causes a serious problem at late-time
evolution for the spherical void model. How do we improve the
description for spherical void evolution?
From the viewpoint of series, the higher order
the perturbation we consider, the narrower the radius
of the convergence becomes.
How do we improve the convergence of the perturbation?

For the contradiction in the Lagrangian approximation,
we consider to improve the convergence rate of a sequence
of partial sums. As a good way to speed up the convergence
of a slowly converging series, Shanks transformation
had been proposed~\cite{Sponier}. First we consider a
simple example. Suppose the $n$-th term in the sequence
takes this form:
\begin{equation} \label{eqn:seq-1}
A_n = A + \alpha q^n ~~(|q|<1) \,.
\end{equation}
The sequence converges $A_n \rightarrow A$
as $n \rightarrow \infty$. To obtain the limit of a
sequence $A$, we solve algebraic equations with
$A_{n-1}, A_n$, and $A_{n+1}$.
\begin{equation}
A = \frac{A_{n+1} A_{n-1} - A_n^2}
 {A_{n+1} + A_{n-1} - 2A_n} \,.
\end{equation}
This formula is exact only if the sequence $A_n$ is
described by the form in (\ref{eqn:seq-1}). For the generic case,
we consider the $n$th term in the sequence takes the form:
\begin{equation}
A_n = A(n) + \alpha q^n \,,
\end{equation}
where for large $n$, $A(n)$ is a more slowly varying function
of $n$ than $A_n$. Let us suppose that $A(n)$ varies sufficiently
slowly so that $A(n-1), A(n)$, and $A(n+1)$ are all approximately
equal.
Then the above discussion motivates the nonlinear transformation
\begin{equation} \label{eqn:Shanks}
S(A_n) = \frac{A_{n+1} A_{n-1} - A_n^2}
 {A_{n+1} + A_{n-1} - 2A_n} \,.
\end{equation}
This transformation is called Shanks transformation,
creating a new sequence $S(A_n)$ which often converges
more rapidly than the old sequence $A_n$.
The sequence $S^2(A_n)= S[S(A_n)]$ and
$S^3(A_n)= S[S[S(A_n)]]$ may be even more
rapidly convergent.

Damour, Jaranowski, and Sch\"{a}fer applied Shanks transformation
to the post-Newtonian approximation of general relativity~\cite{Damour00}.
They improved the analytical determination of various
last stable orbits in circular general relativistic orbits
of two point masses.
In the next section, we apply the transformation to improve
of the Lagrangian perturbation.

%%%%%%%%%%%%%%%%%%%%%%%%%%%%%%%%%%%%%%%%%%%
\section{Improvement of The Lagrangian perturbative solution}\label{sec:example}
%%%%%%%%%%%%%%%%%%%%%%%%%%%%%%%%%%%%%%%%%%%

\begin{figure}
 \includegraphics{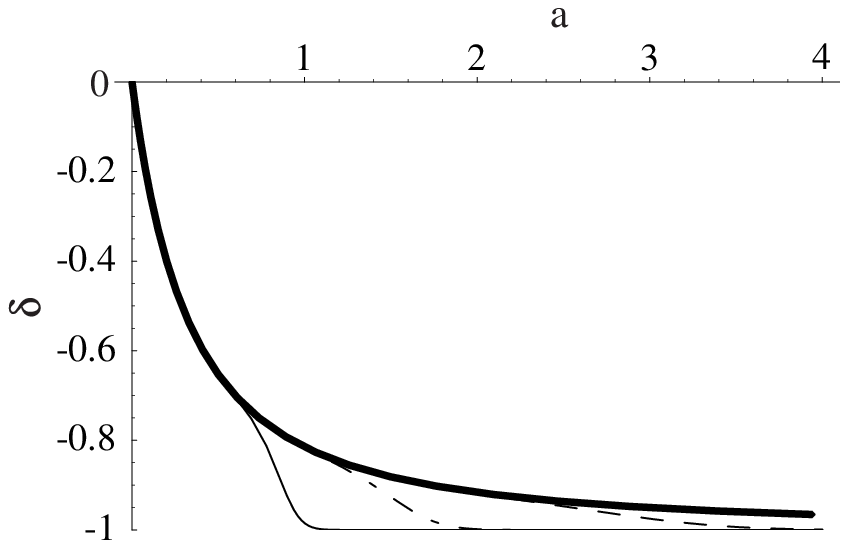}
 \caption{\label{fig:Void2}
 The evolution of the spherical model for negative 
 fluctuation.
 In this figure, the scale factor is normalized as
 in Fig.~\ref{fig:Cluster}.
 We apply Shanks transformation to the Lagrangian perturbative
 solution and improve its accuracy.
 The thick solid line shows the evolution by the exact solution.
 The fine solid line shows the behavior of original eleventh-order
 perturbative solution.
 The dashed-dotted line, and the dashed line
 shows the evolution by once transformed and twice transformed
 perturbative solutions, respectively. 
 After three times transformation, because the difference between exact
 solution and perturbative solution becomes quite small.
 Therefore, if we also draw the curve of three times
 transformed perturbative solutions,
 we cannot find the difference on the figure.
 In this case, 
 by application of Shanks transformation, the Lagrangian approximation
 improves its accuracy. In other words, we can describes
 late-time evolution of voids well. }
\end{figure}

Here we apply Shanks transformation for the Lagrangian perturbation.
We consider the spherical void case and adopt the eleventh-order
solution~\cite{Tatekawa04R}.
From the Lagrangian perturbation (Eq.~(\ref{eqn:Lagrange-void})),
we obtain a new solution via Shanks transformation.
\begin{eqnarray}
\widetilde{R_n} &=& \frac{R_{n+1} R_{n-1} - R_n^2}
 {R_{n+1} + R_{n-1} - 2R_n} \,, \\
R_n & \equiv & \left [1+ \sum_{k=1}^n (-1)^{k+1} C_k a^k \right ] \,.
\end{eqnarray}
From the new sequence or perturbative solution $\widetilde{R_n}$,
we can derive a more refined solution.
Figure~\ref{fig:Void2} shows the evolution of the spherical void
using exact and the Lagrangian perturbative solutions. We
apply Shanks transformation once, twice, and three times
for the perturbation.
After transformation, the solution is refined and recovered
its accuracy. In ordinary Lagrangian perturbation, we cannot
improve the perturbative solution for late-time evolution.
Using Shanks transformation, we can obtain a well-refined
perturbative solution.

%%% Figure %%%
%%%%%%%%%%%%%

We have shown one of the simplest cases. This improvement method
is not limited to a special case. It can be applied
in generic cases.
Suppose the third-order perturbative solution takes this form:
\begin{equation}
S = D_+(t) S^{(1)} + D_+(t)^2 S^{(2)} + D_+(t)^3 S^{(3)} \,,
\end{equation}
where $D_+$ is a linear growing factor.
In the E-dS model, when we consider only the primordial growing
mode in the longitudinal mode, the perturbative solution can
be described by this form exactly~\cite{Buchert94,Bouchet95,Catelan95}.
\begin{eqnarray}
\nabla^2 S^{(2)} &=& -\frac{3}{7} \left [
 (\nabla^2 S^{(1)})^2 - S^{(1)}_{,ij} S^{(1)}_{,ji} \right ] \,,
 \label{eqn:S2} \\
\nabla^2 S^{(3)} &=& -\frac{1}{3} \det \left (S^{(1)}_{,ij}
 \right ) \nonumber \\
 && + \frac{10}{21} \left[ \nabla^2 S^{(1)} \nabla^2 S^{(2)}
 - S^{(1)}_{,jk} S^{(2)}_{,kj} \right ] \,, \label{eqn:S3}
\end{eqnarray}
where $\nabla$ and subscript denote the Lagrangian spacial derivative.
For other universe models, the perturbative solution
can be approximated by this form in the matter-dominant era.
Applying Shanks transformation (\ref{eqn:Shanks}), the
perturbative solution is transformed to
\begin{equation}
\widetilde{S} = D_+ S^{(1)}
 + \frac{D_+^2 (S^{(2)})^2}{D_+ S^{(3)} - S^{(2)}} \,.
\end{equation}
In Sec.~\ref{sec:LCDM}, we will treat a generic case with Shanks
transformation.

%%%%%%%%%%%%%%%%%%%%%%%%%%%%%%%%%%%%%%%%%%%
\section{Comparison with Pad\'{e} approximation}\label{sec:Pade}
%%%%%%%%%%%%%%%%%%%%%%%%%%%%%%%%%%%%%%%%%%%

For a convergence of series, there are other methods. One of these
methods is known as Pad\'{e} approximation~\cite{Sponier}.
Pad\'{e} approximation seems to be a generalization of Taylor
expansion. For a given function $f(t)$, Pad\'{e} approximation
is written as the ratio of two polynomials:
\begin{equation}
f(t) \simeq \frac{\sum_{k=0}^M \alpha_k t^k}{1+ \sum_{k=1}^N \beta_k t^k} \,,
\end{equation}
where $\alpha_k$ and $\beta_k$ are constant coefficients.
Assume we already know the coefficient $\gamma_l$ ($0 \le l \le M+N$)
of the Taylor expansion around $x=0$. Then,
\begin{equation}
f(t) = \sum_{l=0}^{M+N} \gamma_l t^l + o(t^{M+N+1}) \,.
\end{equation}
Comparing the coefficients $\alpha_k$, $\beta_k$, and $\gamma_k$,
we determine $\alpha_k$ and $\beta_k$.
\begin{eqnarray}
\alpha_0 &=& \gamma_0 \,, \\
\alpha_k &=& \sum_{m=1}^N \beta_m \gamma_{k-m} ~~(k=1,\cdots,N) \,, \\
\sum_{m=1}^N \beta_m \gamma_{N-m+k} &=& -\gamma_{N+k} ~~(k=1,\cdots,N) \,.
\end{eqnarray}
The advantage of Pad\'{e} approximation is that
even if we consider a same-order expansion,
Pad\'{e} approximation describes original function
rather better than Taylor expansion does.

Yoshisato, Matsubara, and Morikawa~\cite{Yoshisato98} have proposed
an application of Pad\'{e} approximation for Eulerian perturbative
solutions. Furthermore, Matsubara, Yoshisato, and
Morikawa~\cite{Matsubara98} have applied Pad\'{e} approximation for
the Lagrangian description. They also showed that Pad\'{e}
approximation can improve the Lagrangian perturbative solution. 

\begin{figure}
 \includegraphics{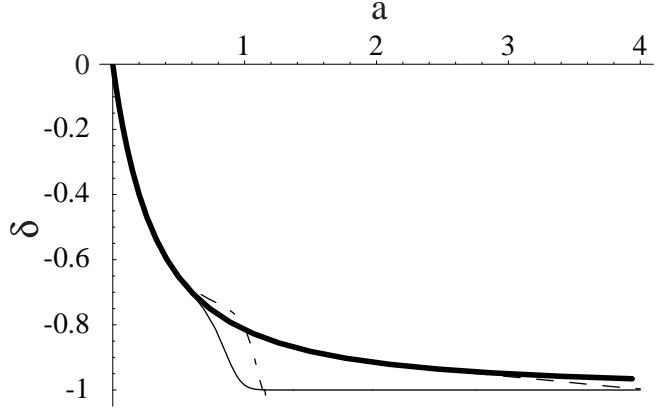}
 \caption{\label{fig:Pade}
 The evolution of the spherical model for negative 
 fluctuation. In this figure, the scale factor is normalized as
 in Fig.~\ref{fig:Cluster}.
 We apply Pad\'{e} approximation to the Lagrangian perturbative
 solution and improve its accuracy.
 The thick solid line shows the evolution by the exact solution.
 The fine solid line shows the behavior of original eleventh-order
 perturbative solution.
 The dashed-dotted line, and the dashed line
 shows the evolution by the case of $(M, N)=(1, 10)$ and $(3, 8)$,
 respectively. When $N$ is greatly different from $M$, 
 the approximation is not improved well. When we choose
 $(M, N)=(5, 6)$, the difference between exact
 solution and perturbative solution becomes quite small.
 Using Pad\'{e} approximation, we can improve the
 perturbative solution well. }
\end{figure}

Here we apply Pad\'{e} approximation to the spherical void.
In Pad\'{e} approximation, it is quite important to choose
the numbers of terms $M$ and $N$.
When $N$ is greatly different from $M$,
the approximation is not improved well.
Fig.~\ref{fig:Pade} shows the evolution of the spherical void
using exact and the Lagrangian perturbative solutions. We
apply Pad\'{e} approximation with several cases.
Here we show the case of $(M, N)=(1, 10), (3, 8)$ and $(5, 6)$.
It is important to choose the parameter $M$ and $N$.
When $N$ is greatly different from $M$, 
not only the approximation is not improved well,
but also the solution will diverges.
For late-time evolution, although the original eleven-order
perturbative solution converges to $\delta \rightarrow -1$,
the Pad\'{e} approximation with $(M, N)=(1, 10), (3, 8)$
diverges. On the other hand, when we choose $(M, N)=(5, 6)$,
the perturbative solution can approximate the exact solution
at late time. We can improve the perturbative solution
with Pad\'{e} approximation, too.
%We will discuss the merits and demerits of Shanks transformation
%in the summary.

%%% Figure %%%
%%%%%%%%%%%%%

%%%%%%%%%%%%%%%%%%%%%%%%%%%%%%%%%%%%%%%%%%%
\section{Generic case: The $\Lambda$CDM model}\label{sec:LCDM}
%%%%%%%%%%%%%%%%%%%%%%%%%%%%%%%%%%%%%%%%%%%

We showed the evolution of a homogeneous spherical void and
noted the improvement for Lagrangian approximation.
Because we know the exact solutions for the spherical collapse
and void evolution, we do not know whether or not the Shanks
transformation is useful for the generic case.
Therefore we must apply the transformation for generic models.

Here we consider a $\Lambda$CDM (Low density Cold Dark Matter) model.
The cosmological parameter at the present time ($z=0$, here we define
$a \equiv 1$ at the present time) is given
by a WMAP 3rd-year result~\cite{WMAP-3rd}:
\begin{eqnarray}
\Omega_m &=& 0.28 \,, \\
\Omega_{DE} &=& 0.72 \,, \\
H_0 &=& 74~ \mbox{[km/s/Mpc]} \,, \\
\sigma_8 &=& 0.74 \,.
\end{eqnarray}
The Gaussian density field is generated by COSMICS~\cite{COSMICS}.
We set up the initial condition at decoupling time ($a=10^{-3}$).
The initial peculiar velocity and the density fluctuation
are adjusted by the growing solution of ZA.

For time evolution, we consider Lagrangian third-order
approximation, Shanks transformation, and N-body simulation.
For computation of the Lagrangian perturbations, we set
the parameters as follows:
\begin{eqnarray*}
\mbox{Number of grids} &:& N=128^3 \,, \\
\mbox{Box size} &:& L=128 h^{-1} \mbox{Mpc}
 ~~(\mbox{at}~a=1)  \,.
\end{eqnarray*}
The N-body simulation is applied by a particle-particle particle-mesh
($P^3M$) method~\cite{P3M} whose code was originally written by
Bertschinger.

For N-body simulations, we set the parameters as follows:
\begin{eqnarray*}
\mbox{Number of particles} &:& N=128^3 \,, \\
\mbox{Box size} &:& L=128 h^{-1} \mbox{Mpc}
 ~~(\mbox{at}~a=1)  \,, \\
\mbox{Softening length} &:& \varepsilon = 50 h^{-1} \mbox{kpc}
 ~~(\mbox{at}~a=1)  \,.
\end{eqnarray*}
Then we impose a periodic boundary condition.

The Lagrangian approximation in $\Lambda$CDM is expanded as
\begin{equation}
S=h_1(t) S^{(1)} + h_2(t) S^{(2)} + h_3(t) S^{(3)} \,,
\end{equation}
where $h_n(t)$ is the growing factor for $n$th-order approximation.
The spacial parts are given by Eqs.(\ref{eqn:S2}) and (\ref{eqn:S3}).
The growing factors are derived with a numerical method~\cite{Bouchet95}.
Applying Shanks transformation~(Eq.(\ref{eqn:Shanks})), the
perturbative solution is transformed to
\begin{equation}
\widetilde{S} = h_1 S^{(1)}
 + \frac{h_2^2 (S^{(2)})^2}{h_3 S^{(3)} - h_2 S^{(2)}} \,.
 \label{eqn:Shanks-3rd}
\end{equation}

In order to avoid the divergence of the density fluctuation,
we need to consider a smoothed density field over the scale $R$.
This density field is related to the unsmoothed density field
$\rho(\bm{x})$ as
\begin{eqnarray}
\rho(\bm{x};R) &=& \int d^3 \bm{y} W(|\bm{x}-\bm{y}|;R) \rho(\bm{y})
\nonumber\\
&=& \int \frac{d^3 \bm{k}}{(2\pi)^3} \tilde{W} (kR)
\tilde {\rho} (\bm{k}) e^{-i \bm{k}\cdot \bm{x}} \,,
\end{eqnarray}
where $W$ denotes the window function and
$\tilde{W}$ and $\tilde{\rho}$ represent the Fourier transforms
of the corresponding quantities.
In this paper, we adopt the top hat window function,
\begin{eqnarray}
\tilde{W} = \frac{3(\sin x - x \cos x)}{x^3}\,.
\end{eqnarray}
Then, the density fluctuation $\delta(\bm{x};R)$
at the position $\bm{x}$ smoothed over the scale $R$ can be constructed
in terms of $\rho(\bm{x};R)$.

Here we choose the smoothing scale $R=1 h^{-1} \mbox{Mpc}$. Then
we calculate the power spectrum of density fields.
In order to obtain the power spectrum,
we generate 50 samples for the primordial density
fluctuations. Then we pick up snapshots at $z=5, 4, 3, 2, 1, 0$.

%%% Figure %%%
\begin{figure}
 \includegraphics{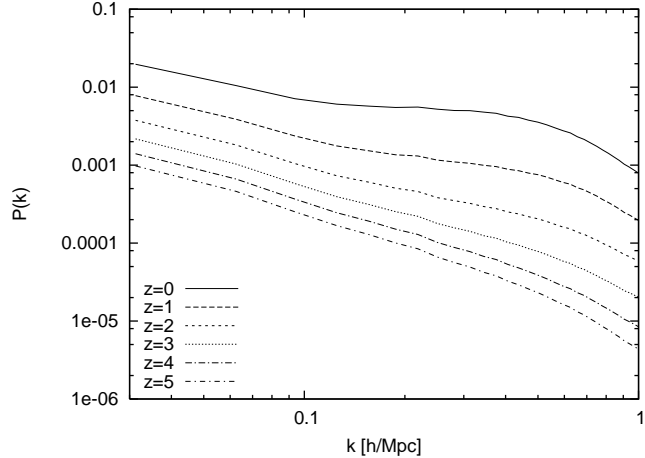}
 \caption{\label{fig:LCDM-P3M}
 The power spectrum of the density
 field in N-body simulation. 
 Because of a nonlinear effect in small structures,
 the large-$k$ components in the power spectrum 
 grow remarkably.
 }
\end{figure}
%%%%%%%%%%%%%

Figure~\ref{fig:LCDM-P3M} shows the power spectrum of the density
field in N-body simulation. During evolution, because a strongly
nonlinear region promotes the growth of the fluctuation,
the large-$k$ components in the power spectrum grows remarkably.

%%% Figure %%%
\begin{figure}
 \includegraphics{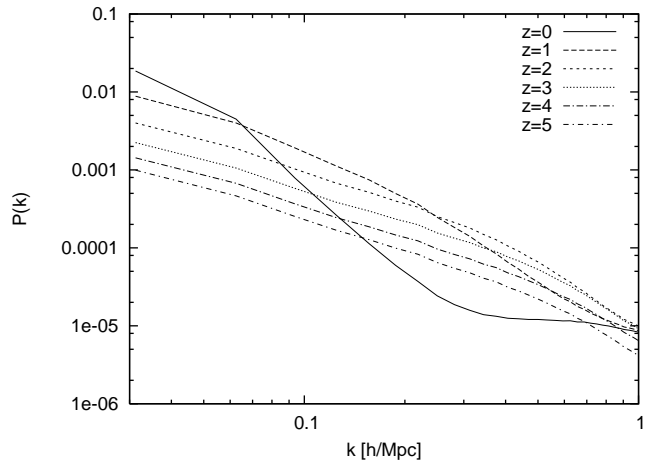}
 \caption{\label{fig:LCDM-LA3}
 The power spectrum of the density
 field in Lagrangian third-order approximation. 
 Because of the shell-crossing,
 the spectrum distorted from the large-$k$ components
 to the small-$k$ components gradually.
 At $z=0$, the spectrum is really differs greatly
 that in N-body simulation. 
}
\end{figure}
%%%%%%%%%%%%%

On the other hand, the power spectrum in the Lagrangian third-order
spectrum sinks (Fig.~\ref{fig:LCDM-LA3}). 
At the high-$z$ era, shell-crossing occurs in small
structures. After shell-crossing, the cluster it forms
spreads eternally. Then the negative influence of shell-crossing
affects the large-scale structure. Therefore the spectrum sinks
from the small scale gradually. At $z=0$, the spectrum differs
greatly from that in N-body simulation. 

%%% Figure %%%
\begin{figure}
 \includegraphics{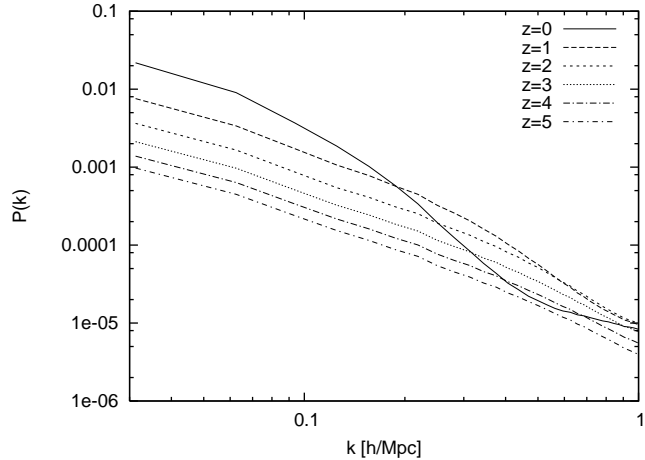}
 \caption{\label{fig:LCDM-SA3}
 The power spectrum of the density
 field in Shanks transformation. 
 Because of the improvement of the nonlinear effect,
 the distortion of the spectrum is improved well. 
}
\end{figure}
%%%%%%%%%%%%%

When we apply Shanks transformation, although a strongly nonlinear
effect in the large-$k$ components is not realized, the distortion
of the spectrum is well improved~(Fig.~\ref{fig:LCDM-SA3}).
Even if we consider a low-$z$ era, Shanks transformation
can realize the spectrum, except for the small scale, well.

%%%%%%%%%%%%%%%%%%%%%%%%%%%%%%%%%%%%%%%%%%%
\section{summary}\label{sec:summary}
%%%%%%%%%%%%%%%%%%%%%%%%%%%%%%%%%%%%%%%%%%%

We have discussed the evolution of the spherical void in
the framework of the Lagrangian perturbation.
Using ordinary Lagrangian perturbation,
the higher-order Lagrangian approximation
deviates from the exact solution at late time.
In other words, ZA gives the best description for the late-time
evolution of voids.
Then we generalized Shanks transformation for the Lagrangian
perturbation, i.e., we analyzed time evolution for a $\Lambda$CDM
model.

We apply Shanks transformation, which accelerates
the convergence of series for the Lagrangian perturbation.
Although the transformation in valid within a convergent radius
of the series, the transformation creates a new sequence
which often converges more rapidly than the old sequence.
In the spherical void model, the transformation is valid for
a long time. Then we can improve the accuracy of the Lagrangian
description. Using this method, we can solve the problem
whereby the higher order the perturbation we consider, the worse
the approximation becomes in late-time evolution.

In this paper, we also compare the accuracy of our improved methods,
between Shanks transformation and  Pad\'{e} approximation, that is.
In the comparison, we found that
Shanks transformation has several merits: 
both Shanks transformation and Pad\'{e} approximation
can be derived with algebraic procedures.
In Pad\'{e} approximation, although we can solve algebraic equations
and find unique solutions, the equation is quite complicated.
Furthermore, according to our analyses, if the difference
between two parameters $M$ and $N$ in Pad\'{e} approximation
is large, the perturbative solution will diverge.
On the other hand, Shanks transformation does not
diverge.

However, in several points, Shanks transformation
shows its weakness.
To apply Pad\'{e} approximation, we need to obtain
second-order perturbation. On the other hand, when
we consider the Shanks transformation, we must obtain
at least third-order perturbation. 
Using the Shanks transformation, we obtain a new perturbative
solution $\widetilde{R}_n$ from $R_{n-1}, R_{n}, R_{n+1}$.
Then, to repeatedly apply Shanks transformation,
we require $R_{n-2}, \cdots, R_{n+2}$.
In general, when we apply $n$ times transformation,
at least we must know $2n+1$-th order perturbative solutions.
To improve the perturbation well, we must repeat
the transformation several times.

From the viewpoint of algebraic procedures, Shanks
transformation has an advantage. In Pad\'{e} approximation,
we must solve nonlinear simultaneous equations. Then the
solution is extremely complicated in a higher-order case.
For example, when we improve an eleventh-order solution
with $(M, N)=(5, 6)$, we derived about fifty-digit
coefficients.

For a $\Lambda$CDM model, Shanks transformation recovers
accuracy for the description of the density field, too.
We showed that
the power spectrum in Shanks transformation
becomes better than that of ordinary Lagrangian approximations.
Even if we consider the spectrum at $z=0$, the Shanks
transformation can describe quite well,
except for the small scale.

However, in a generic case, the critical problem 
in Shanks transformation appeared.
When we continue to apply Shanks transformation, the
divergence of the perturbation occurs. For example,
when we apply the transformation with third-order approximation,
the perturbation is written as Eq.~(\ref{eqn:Shanks-3rd}).
If the higher-order perturbation is smaller than
the lower-order perturbation, we consider that the perturbative
method is valid. From Eq.~(\ref{eqn:Shanks-3rd}), when
second-order and third-order perturbation become
\begin{equation}
h_2 S^{(2)} \simeq h_3 S^{(3)} \,,
\end{equation}
the perturbation diverges. In other words, because
the Lagrangian perturbation is excluded from the convergence
series during evolution, the perturbative method
becomes invalid. Therefore when we apply Shanks transformation,
we notice the validity of the perturbative expansion.
In the spherical collapse case, the density fluctuation diverges
before divergence of the Lagrangian perturbation.
After the shell-crossing, the divergence of the perturbation
is settled. Then the perturbation describes the displacement
of the fluid.

Even if we improve the Lagrangian perturbation, we cannot avoid
the problem of shell-crossing. Because of shell-crossing, the
approximation becomes worse at late time. To solve this
problem, Scoccimarro and Sheth proposed an extrapolation method
~\cite{Scoccimarro02}.
First, they considered structure formation with
second-order Lagrangian approximation. After that,
they extrapolated the density distribution in the high-density
region from NFW profile. This hybrid method realizes the density
distribution well. Although it is very complicated,
this method is effective for higher-order perturbation.
We can then expect that this hybrid method will realize
the density distribution with high precision by
combining it with third-order Lagrangian approximation or
Shanks transformation.

\begin{acknowledgments}
We thank Masahiro Morikawa and Akira Ohashi
for useful discussion.
We thank Tomohiro Harada and Atsushi Taruya
for their useful comment.
We would also like to thank Peter Musolf for checking
our paper's English. 
The numerical calculations were partially carried out
on Altix3700 BX2 at YITP, Kyoto University.
\end{acknowledgments}

\end{document}